\documentclass[acus]{JAC2003}

\usepackage{graphicx}
\usepackage{booktabs}
\usepackage{amsmath}
\usepackage{amssymb}

\def\beq{\begin{equation}}
\def\eeq{\end{equation}}
\def\bdis{\begin{displaymath}}
\def\edis{\end{displaymath}}

\def\case#1/#2{\textstyle\frac{#1}{#2}}

\begin{document}



\onecolumn
\begin{flushright}
\large
FERMILAB-CONF-10-410-AD-APC\\[1in]
\end{flushright}
\begin{center}
{\LARGE 
APPLICATION OF A LOCALIZED CHAOS\\[0.06in]
BY RF-PHASE MODULATIONS\\[0.16in] 
IN PHASE-SPACE DILUTION}
\\[0.4in]
\large
S.Y.~Lee\\[0.2in]
{\em Indiana University, Bloomington, IN 47405}\\[0.2in]
K.Y.~Ng\\[0.2in]
{\em Fermilab, Batavia, IL 60510}\\[0.4in]
\begin{abstract}
\parbox{5in}{
Physics of chaos in a localized phase-space region is exploited to produce 
a longitudinally uniformly distributed beam. Theoretical study and 
simulations are used to study its origin and applicability in phase-space 
dilution of beam bunch. Through phase modulation to a double-rf system, 
a central region of localized chaos bounded by invariant tori are generated 
by overlapping parametric resonances. 
Condition and stability of the chaos 
will be analyzed.  
Applications 
include high-power beam, beam distribution uniformization, and industrial 
beam irradiation.} 
\end{abstract}
\vskip0.5in
Submitted to\\
46th ICFA Advanced Beam Dynamics Workshop on\\ 
High-Intensity and High-Brightness Hadron Beams (HB2010)\\
Morschach, Switzerland\\
September 27--October 1, 2010
\end{center}
\newpage


\title{APPLICATION OF A LOCALIZED CHAOS GENERATED BY RF-PHASE
MODULATIONS IN PHASE-SPACE DILUTION\thanks{Work supported by 
the US DOE under contracts DE-FG02-92ER40747,
DE-AC02-76CH030000, and 
the NSF under contract NSF PHY-0852368.
}\vspace{-0.18in}}

\author{S.Y. Lee, Indiana University, Bloomington, IN 47405, US\\ 
K.Y. Ng, Fermilab, Batavia, IL 60510, US}

\maketitle
\setlength{\abovedisplayskip}{4pt plus 0pt minus 3pt}
\setlength{\belowdisplayskip}{4pt plus 0pt minus 3pt}

\begin{abstract}\vskip-0.04in
Physics of chaos in a localized phase-space region is exploited to produce 
a longitudinally uniformly distributed beam. Theoretical study and 
simulations are used to study its origin and applicability in phase-space 
dilution of beam bunch. Through phase modulation to a double-rf system, 
a central region of localized chaos bounded by invariant tori are generated 
by overlapping parametric resonances. 
Condition and stability of the chaos 
will be analyzed.  
Applications 
include high-power beam, beam distribution uniformization, and industrial 
beam irradiation.\\[-0.19in]  
\end{abstract}

\section{INTRODUCTION\vspace{-0.03in}}

ALPHA, under construction at IU CEEM, is  
a 20-m electron storage ring.~\cite{alpha}
The project calls for storing a tiny 
synchrotron-radiation-damped
bunch to be extended
to about 40~ns with 
uniform longitudinal distribution.
RF barriers should be the best candidate for bunch lengthening.
Unfortunately, this ring is only 66.6~ns in length, and the widths of the
barriers must be of the order of 10~ns or less.  The risetime of the
barrier voltage will therefore be 
a few ns, or the rf
generating the barrier voltage will be in the frequency range of 
a few hundred MHz.  Ferrite is very lossy at such high frequencies 
and is therefore unsuitable for the job.  
Even if another material could be substituted,
the barriers of such narrow widths would require very high rf peak voltage;
the rf system would be very costly.

Another way to achieve bunch lengthening is to perform phase modulation of
the rf wave so as to produce a large chaotic region at the center of the
rf bucket, but bounded by well-behaved tori.  
The 
beam at the bucket center will be blown up to the much larger
chaotic region.  If true chaoticity is achieved, the particle distribution
will be uniform.  Such an idea has been demonstrated experimentally at the
IUCF Cooler ring in 1997,~\cite{jeon}  
where a double-rf system was used
and the diffusion was found rather sensitive to the 
phase difference $\Delta\phi_0$ between the two rf waves.  
In this paper, the modulation method is further investigated by first
determining the choice of $\Delta\phi_0$, and next analyzing the condition
and stability of the localized chaotic \vspace{-0.04in}region.

\section{THE MODEL\vspace{-0.04in}}
The model to be studied is described by the Hamiltonian
$H=H_0+H_1$, where~\cite{liu}
\bdis
H_0\!=\frac12\nu_s\delta^2
\!+\!\nu_s\left\{1\!-\!\cos\phi
\!-\!\frac{r}{h}\big[1\!-\!\cos(h\phi\!+\!\Delta\phi_0)\big]\right\}\!,
\edis
\beq
H_1=a\delta\nu_s\sin(\nu_m\theta+\eta).
\label{h1}
\eeq
Here $r$ is the ratio of the two rf voltages, 
$h$ is the ratio of the two rf harmonics, 
$\nu_s$ is the small-amplitude synchrotron tune in
the absence of the second rf, 
$\nu_m$ is the phase modulation tune, 
$\eta$ is the modulation phase, 
$a$ is the modulation amplitude, 
$\phi$ is the rf phase, 
$\delta$ is the canonical momentum offset, and 
$\theta$ advances by $2\pi$ per revolution turn.
This model entails a number of parameters.  In this paper, however,
we restrict ourselves to the special case of $r=1/2$, $h=2$,
$\nu_m/\nu_s=2$, and $\eta=0$, thus leaving behind only the phase offset
$\Delta\phi_0$ and the modulation amplitude \vspace{-0.06in}$a$.

\subsection{Choice of $\Delta\phi_0$\vspace{-0.05in}}
The action at the bottom of an rf potential well $V(\phi)$ 
is zero and so are the
{\em resonance strengths\/} generated by phase modulation (see  
Fig.~\ref{strength-30deg-fig} below).  Since the bunch will be tiny,
it will be difficult to be driven into parametric resonances if
it sits at the bottom of the rf potential.  This explains why a
two-rf system is necessary.  The phase difference $\Delta\phi_0$ between
the two rf's shifts the potential-well bottom away from the center
of the longitudinal phase space, where the tiny bunch is located.
The action at the bunch is now finite.
Thus the farther the potential-well bottom is shifted, the larger 
the resonant strengths.

\begin{figure}[b]
\vspace{-0.2in}
\parbox{0.6in}{\vspace{-0.3in}
\caption{\small 
\mbox{Offset of} potential-well bottom 
\mbox{$\phi_0$ as a} 
\mbox{function of} 
\mbox{rf phase} difference $\Delta\phi_0$ at $h=1/r=2$.
\label{minimum-fig}}
}~~
\parbox{2.4in}{
\includegraphics[angle=-90,width=2.45in,viewport=75 10 560 690,clip]
 {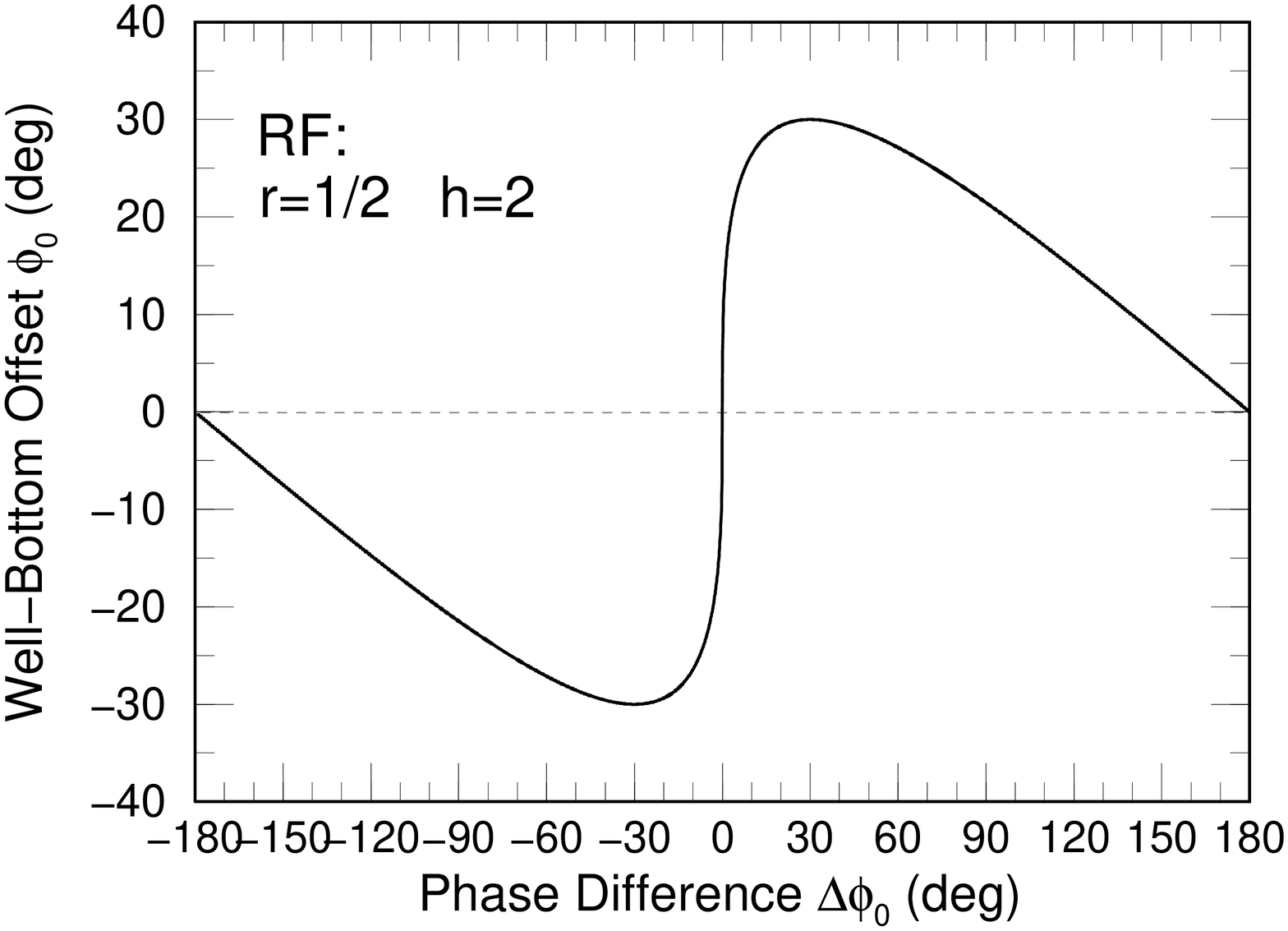}}
\end{figure}

To generate a large region of chaoticity, 
the eventual modulation amplitude $a$ will be large. However,
a perturbative approach is taken in the analysis so as to get a ball-park
understanding of the mechanism.
For the unperturbed Hamiltonian, the position of the 
potential-well bottom $\phi_0$
is given by
$V'(\phi_0)/\nu_s=
\sin\phi_0-r\sin(h\phi_0+\Delta\phi_0)=0,$
and is at a maximum when $V''(\phi_0)=0$.  This leads to
the solution $\phi_0=\pm\sin^{-1}r$.  The corresponding phase 
difference  between the two rf's is therefore (see Fig.~\ref{minimum-fig})
\beq
\Delta\phi_0=\frac{\pi}{2}-h\sin^{-1}r.
\eeq
When $h\!=\!1/r\!=\!2$, the largest offset of well bottom is
$\phi_0\!=\!\pm30^\circ$ and the corresponding phase difference between the two
rf's is $\Delta\phi_0=\pm30^\circ$.
Simulations show that diffusion of a tiny bunch at the phase-space center
is possible when $20^\circ\lesssim|\Delta\phi_0|\lesssim50^\circ$.
In below, we first study the case of $\Delta\phi_0=30^\circ$, and attempt 
another value later.

\subsection{Synchrotron Tune and Resonance Strengths\vspace{-0.02in}}
\begin{figure}[b]
\vspace{-0.18in}
\includegraphics[angle=-90,width=3.1in,viewport=75 10 560 683,clip]
{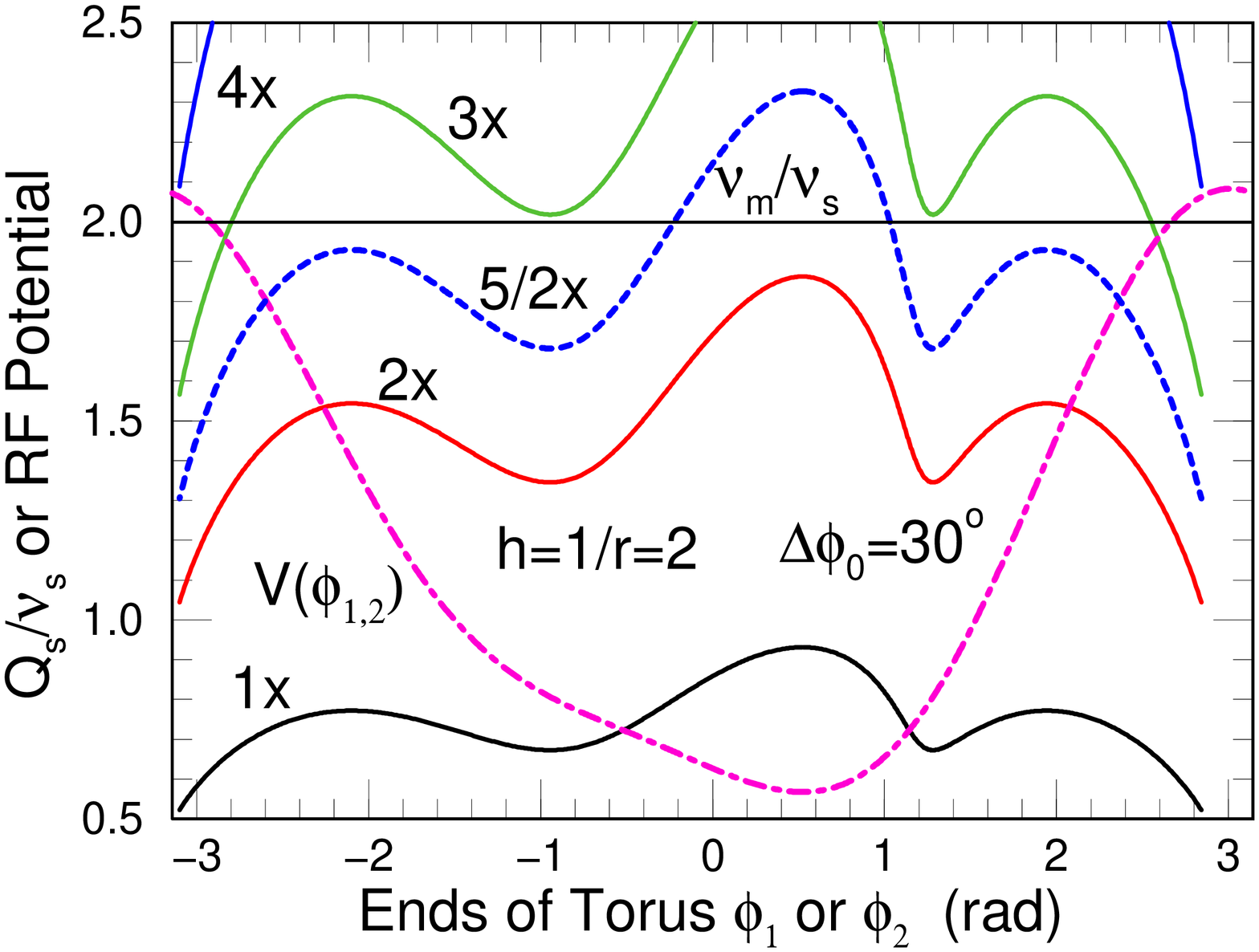}
\vspace{-0.1in}
\caption{\small (color)
Synchrotron tune $Q_s/\nu_s$ versus  oscillating
torus smaller end $\phi_1$ or larger end $\phi_2$, 
with $h\!=\!1/r\!=\!2$ and $\Delta\phi_0\!=\!30^\circ$.  
Some multiples are also plotted
to illustrate the possible parametric resonances driven the phase
modulation tune $\nu_m/\nu_s\!=\!2$ (horizontal line.) 
The asymmetric rf potential is shown in
dot-dashes.  
\label{syntune-30deg-fig}}
\vspace{-0.08in}
\includegraphics[angle=-90,width=3.1in,viewport=75 10 555 683,clip]
{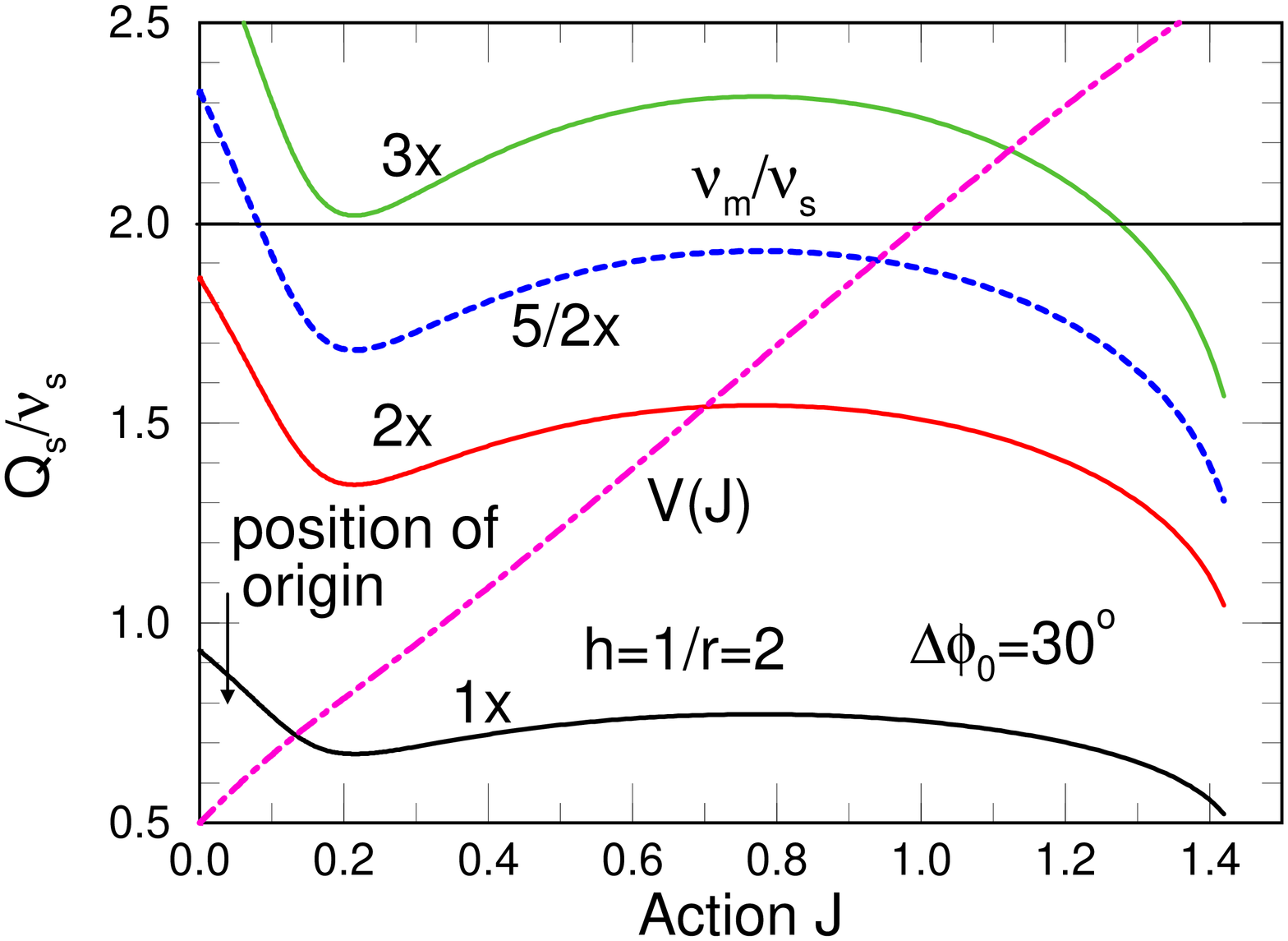}
\vspace{-0.12in}
\caption{\small (color) Synchrotron tune $Q_s/\nu_s$ and its multiples 
as functions of action $J$,
with $h=1/r=2$ and $\Delta\phi=30^\circ$.  
The rf potential $V(J)$ is shown in dot-dashes.
\label{syntune-J-30deg-fig}}
\end{figure}

For the unperturbed Hamiltonian, the action and angle of an oscillatory orbit
are given by
\beq
J=\frac{1}{2\pi}\oint\delta(\phi)d\phi,
~~~\psi
=\frac{Q_s}{\nu_s}\int_{\phi_1}^\phi\frac{d\phi'}{\delta(\phi')},
\label{psi}
\eeq 
where $\phi_1$ is smaller end of the phase excursion. 
When $\phi$ reaches the larger end of the phase excursion 
$\phi=\phi_2$ while
$\psi$ advances by $\pi$,  
the synchrotron tune $Q_s$ of the oscillatory torus can 
be extracted, and  
is depicted
in Fig.~\ref{syntune-30deg-fig}, where the asymmetric rf potential is also
plotted in dot-dashes.
With the modulation tune $\nu_m/\nu_s\!=\!2$ (horizontal line), we
see that the tiny bunch at the phase-space center will be driven 
into 3:1 and 5:2 resonances.
The synchrotron tune can also be computed as a function of the action $J$,
and this is depicted in Fig.~\ref{syntune-J-30deg-fig}.
It is important to point out that we require the synchrotron tune to be large
at the central part than the edges of the phase space, because we wish to
have a central chaotic region bounded by well-behaved tori.  

\begin{figure}[h]
\vspace{-0.1in}
\includegraphics[angle=-90,width=3.05in,viewport=75 10 560 686,clip]
{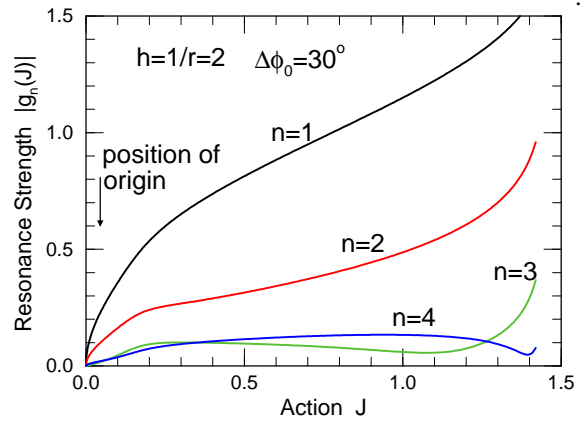}.
\vspace{-0.15in}
\caption{\small (color)
Resonance strengths $|g_n(J)|$ as functions of the action $J$ for 
$n\!:\!1$  parametric resonances, with $n\!\leq\!4$, 
$h\!=\!1/r\!=\!2$.
\label{strength-30deg-fig}}
\vspace{-0.28in}
\end{figure}

To express the perturbative Hamiltonian $H_1$ of Eq.~(\ref{h1})
in terms of action-angle variables, the reduced momentum offset
$\delta$ is expanded into Fourier series,\vspace{-0.04in}
\beq
\delta\!=\!\sum_{n=-\infty}^\infty\! g_n(J)e^{in\psi},~~~
g_n(J)\!=\!\frac{1}{2\pi}\!\int_{-\pi}^\pi\delta  e^{-in\psi}d\psi.
\eeq
Keeping only the first-order perturbative terms, the Hamiltonian can
now be expressed as
\begin{align}
H\!=&H_0(J)+a\nu_s\bigg\{\!\sum_{n>0}|g_n(J)|
\Big[\sin(\nu_m\theta\!+\!\eta\!+\!\chi_n\!+\!n\psi)~~~~~~\nonumber\\
&\!\!\!\!+\sin(\nu_m\theta\!+\!\eta\!+\!\chi_n\!-\!n\psi)\Big]
\!+g_0(J)\sin(\nu_m\theta\!+\!\eta)\!\bigg\},\!\!\!\!\!
\label{first-order}
\end{align}
with $\chi_n$ the phase of $g_n$,
showing all the first-order parametric resonances.
The resonance strength function $|g_n|$ is a measure of its ability to drive
the $n\!:\!1$ parametric resonance, and is depicted
in Fig.~\ref{strength-30deg-fig} for $n\!=\!1$ to 4.
Note that all the strength functions vanish at $J=0$, which is the
reason why we need to shift the potential-well bottom away from the
center of the phase space.  Because of the asymmetry
of $\delta(\phi)$, $g_n(J)$ no longer
vanishes when $n$ is even, implying that both 2:1 and 3:1 resonances
can be excited depending on the choice of the 
modulation tune $\nu_m$.

\section{SIMULATIONS WITH $\Delta\phi_0=30^\circ$\vspace{-0.03in}}

Simulation is performed by tracking each macro-particle
from the $k$-th turn to the $(k\!+\!1)$-th turn
according to the Hamiltonian of Eq.~(\ref{h1}):
\begin{align}
\!\!\!\phi_{k+1}&\!=\!\phi_k\!+\!2\pi\nu_s\big[\delta_k
\!+\!a\sin(2\pi k\nu_m )\big],  \nonumber\\
\!\!\!\delta_{k+1}&\!=\!\delta_k\!-\!2\pi\nu_s\big[\sin\phi_{k+1}
\!-\!r\sin(h\phi_{k+1}+\Delta\phi_0)
\big],
\end{align}
where the assumption of slowly varying particle position within a turn
has been applied.  The modulation period has been 
chosen to be exactly 515 turns
with $\nu_m/\nu_s=2$.  
The modulation phase has been taken to be $\eta=0$.
Other values of $\eta$ will lead to different rotated
stroboscopic views of the tracking results. However, all the conclusions
on diffusion and beam enlargement will not be affected.

\begin{figure}
\includegraphics[angle=-90,width=3.1in,viewport=75 10 560 686,clip]
{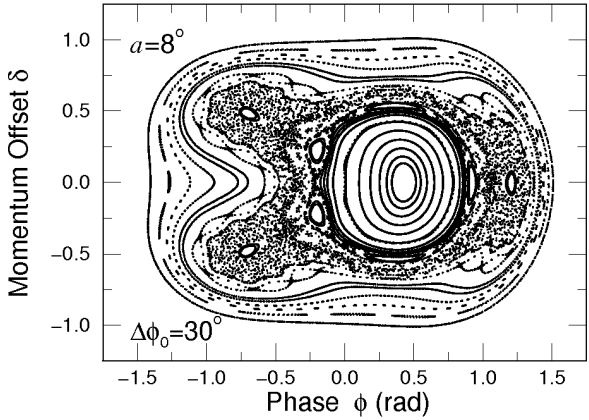}
\vspace{-0.1in}
\caption{\small 
Structure of phase space at $a=8^\circ$ and
$\Delta\phi_0=30^\circ$.
\label{a8-30deg-8-fig}}
\vspace{-0.10in}   
\end{figure}

We first study
the structure of the phase space at modulation amplitude $a=8^\circ$ and
$\Delta\phi_0=30^\circ$, as depicted in Fig.~\ref{a8-30deg-8-fig}. 
This is the cumulative stroboscopic modulation-period views 
in half million turns.
The central stable region is bounded by the 5:2 resonance.  After that
we see the remnant of the 8:3 resonance which merges partially with
the 3:1 resonance.  Then comes the 25:8 resonance and the well-behaved tori.
In order for the bunch, initially at the phase-space center, to 
be enlarged via diffusion, we require the 5:2 resonance to collapse
and the central stable region to shrink so that the bunch is inside the
chaotic region initially.  This occurs when the modulation
amplitude increases to $a=46^\circ$.

\begin{figure}
\includegraphics[angle=-90,width=3.1in,viewport=75 10 570 683,clip]
{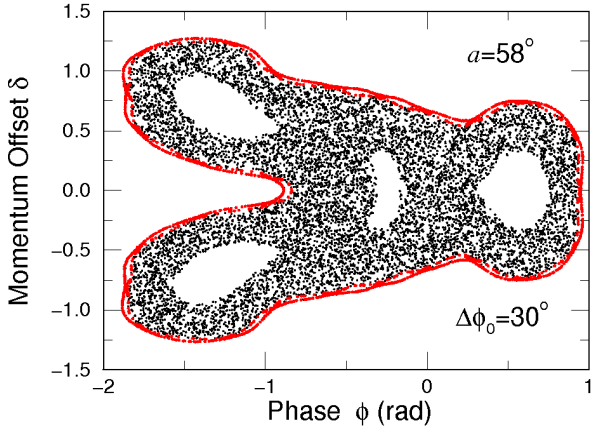}
\vspace{-0.1in}
\caption{\small (color) 
Black: Diffusion of 
a tiny Gaussian bunch of rms spread $\sigma_\phi=0.001$~rad at 
phase-space center at the last modulation period
in 1.2 million turns.  Modulation amplitude is
$a=58^\circ$.   
Red: Stroboscopic plot of one particle initially at $\phi=0.94$ rad
and $\delta=0$ at every modulation period.
\label{a58-30deg-82-fig}}
\vspace{-0.15in}
\end{figure}

\begin{figure}
\includegraphics[angle=-90,width=3.1in,viewport=75 10 560 686,clip]
{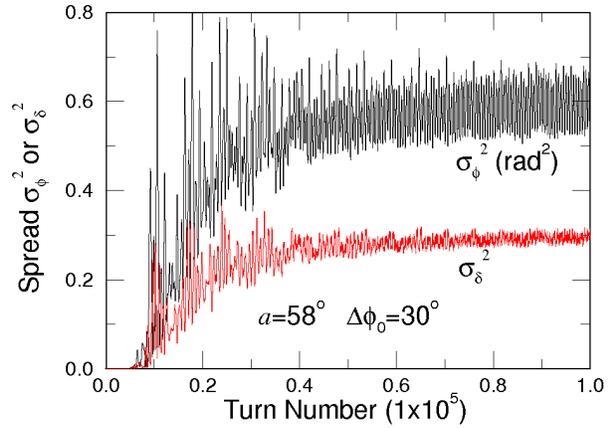}
\vspace{-0.08in}
\caption{\small (color)
Longitudinal bunch size versus turn number
at $a=58^\circ$ and $\Delta\phi_0=30^\circ$ as diffusion develops.
\label{a58-30deg-93-fig}}
\vspace{-0.2in}
\end{figure}

Next
a Gaussian distributed bunch of rms spread $\sigma_\phi=0.001$~rad
consisting of 10000 macro-particles 
at the center of the longitudinal phase space is tracked.    
The particle distribution with
modulation amplitude $a=58^\circ$ is
shown in Fig.~\ref{a58-30deg-82-fig} 
at the last modulation
period in 1.2 million turns.  Essentially, the tiny
bunch at the phase-space center is driven into the thick stochastic layers
surrounding the separatrices of the 3:1 resonance.
The thick stochastic layers, on the other hand, come from the overlapping of 
the 5:2, 8:3, and possibly many other higher-order resonances, which
are not included
in the first order perturbation of the modulation 
presented in Eq.~(\ref{first-order}).
The distribution appears to be uniform except for the four big empty space,
where the four stable fixed points of the 3:1 resonance are located.
The rms beam size is computed turn by turn and is depicted in 
Fig.~\ref{a58-30deg-93-fig}.
We see that the rms-bunch-size squared, $\sigma_\phi^2$ or $\sigma_\delta^2$, 
grows linearly with turn number, signaling that the bunch enlargement 
is indeed a diffusion process.
The growth levels off after about
$1\times10^5$ to $2\times10^5$ turns.
The rms phase and momentum spreads increase to $\sigma_\phi=0.81\pm0.03$~rad 
and $\sigma_\delta=0.57\pm0.01$, respectively.

We next continue the tracking by doubling the number of turns; the distribution
remains bounded and the pattern does not change.  We also track a particle
initially at $\phi\!=\!0.94$~rad and $\delta\!=\!0$ for $1\!\times\!10^5$ turns.  
Its
positions at every modulation period are shown as red dots in 
Fig.~\ref{a58-30deg-82-fig}.  These red dots constitute a well-behaved 
chain of islands, 
confirming that
the diffused bunch will be well-bounded.\vspace{-0.05in}

\subsection{Variation of Modulation Amplitude\vspace{-0.05in}}

\begin{figure}[b]
\vspace{-0.15in}
\includegraphics[angle=-90,width=3.1in,viewport=75 10 560 686,clip]
{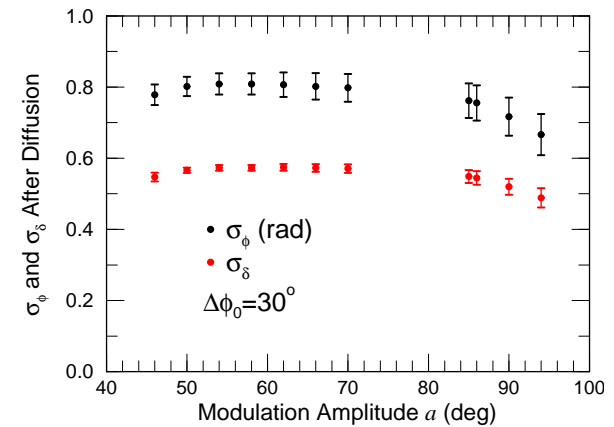}
\vspace{-0.1in}
\caption{\small (color) 
Bunch spreads after diffusion at various modulation amplitudes 
with $\Delta\phi_0=30^\circ$.
\label{bunchspread-30deg-fig}}
\end{figure}

When the modulation amplitude is varied from $a=46^\circ$ to $70^\circ$,
there is not much difference in the shape of the final diffused
bunch distribution.   The only significant change is the gradual left-shifting
of the chaotic pattern in Fig.~\ref{a58-30deg-82-fig},
which is a consequence of the detuning of the
synchrotron tune $Q_s$ as the modulation amplitude $a$ increases.
The diffused rms phase and momentum spreads are still roughly 
at $\sigma_\phi\approx0.8$~rad and $\sigma_\delta\approx0.58$,
respectively, as depicted in Fig.~\ref{bunchspread-30deg-fig}.
When the modulation strength is increased past $a=70^\circ$, suddenly
no diffusion is observed independent of how long the simulations are
performed.  This turns out to be the moment when the rightmost empty region 
in Fig.~\ref{a58-30deg-82-fig} has
been left-shifted to include the phase-space origin.
The bunch is now inside the rightmost island of the 3:1 resonance making 
small tori around the stable fixed points of
the 3:1 resonance.
 
The diffusion of the bunch does return
when the modulation strength increases up to $a\geq85^\circ$.
Now the three outside islands of the 3:1 resonance are completely filled up
by higher-order resonances so that the tiny bunch initially at the phase-space
center can diffuse outward again.

\begin{figure}
\includegraphics[angle=-90,width=3.1in,viewport=75 10 560 686,clip]
{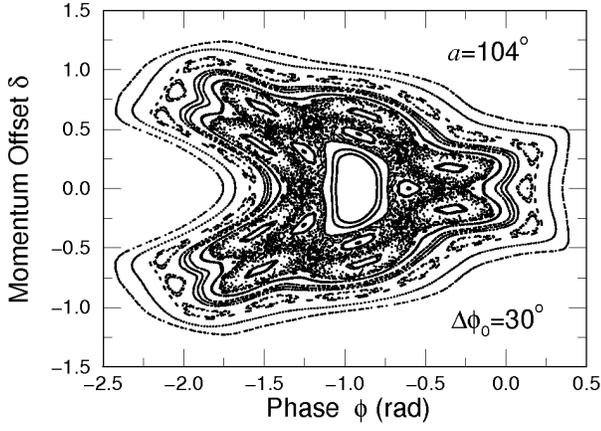}
\vspace{-0.1in}
\caption{\small Stroboscopic plot of phase-space structure at $a=104^\circ$
and $\Delta\phi=30^\circ$,
showing that the bunch original position lies on a well-behaved torus
completely outside the chaotic region.
\label{a104-30deg-8-fig}}
\vspace{-0.2in}
\end{figure}

As the modulation amplitude continues to increase, the phase-space structure
tends to contract and shift further to the left.
As $a>94^\circ$, the bunch original position moves out of the
chaotic region and no diffusion occurs.  A typical phase-space pattern 
at $a=104^\circ$ is shown in Fig.~\ref{a104-30deg-8-fig}.
As $a$ increases past $110^\circ$, beam loss occurs.

\section{SIMULATIONS WITH $\Delta\phi_0=45^\circ$}

Figures~\ref{syntune-30deg-fig} and \ref{syntune-J-30deg-fig} shows
that the approximate intercepts of horizontal line $\nu_m/\nu_s=2$
with the $3\times Q_s/\nu_s$ curve are far away from the initial location
of the particle bunch.  In other words, the bunch is initially far from
the unstable fixed points of the 3:1 resonance.  
This limits the bunch from falling inside the stochastic layers 
surrounding the separatrices unless the modulation
amplitude is sufficiently large.  Figure~\ref{strength-30deg-fig} shows that
the strength function  of the 3:1 resonance is much smaller than that of the 
2:1 resonance and as a result very large modulation amplitude has to be
employed.  All these reasons educe us to a deviation 
from the maximum potential-well-bottom offset.
Here, we try the rf phase difference $\Delta\phi_0\!=\!45^\circ$, which leads
to a well-bottom offset of $\phi_0\!=\!29.12^\circ$, which is only 3\%
less than the maximum value of $30^\circ$.  
This explains why the range
of $\Delta\phi_0$ that  can produce bunch lengthening is not too narrow.

\begin{figure}
\includegraphics[angle=-90,width=3.1in,viewport=75 10 560 683,clip]
{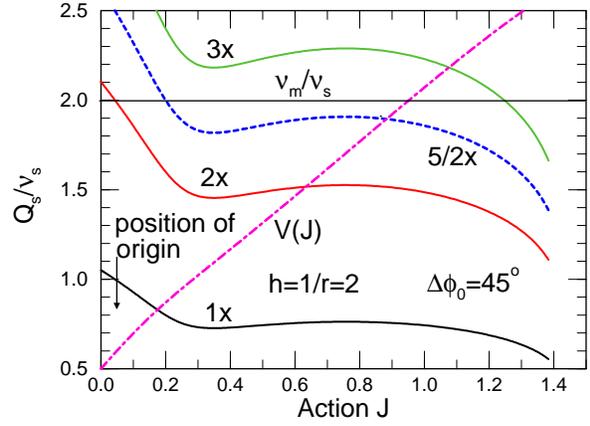}
\vspace{-0.1in}
\caption{\small (color)
Synchrotron tune $Q_s/\nu_s$ as a function of action $J$,
with $h=1/r=2$ and $\Delta\phi_0=45^\circ$.
$2\times$, $5/2\times$, and $3\!\times\! Q_s/\nu_s$ are also plotted
to illustrate the possible parametric resonances driven at the phase
modulation frequency $\nu_m/\nu_s=2$. 
The rf potential $V(J)$ is shown in dot-dashes.
\label{syntune-J-45deg-fig}}
\vspace{-0.1in}
\end{figure}

Corresponding to Fig.~\ref{syntune-J-30deg-fig}, the synchrotron tune
as a function of action for $\Delta\phi_0=45^\circ$ is shown in 
Fig.~\ref{syntune-J-45deg-fig}.
Observe that where the horizontal line $\nu_m/\nu_s\!=\!2$ cuts the
$2\times Q_s/\nu_s$ curve is extremely close to  $J\!=\!0.053$, the
initial location of bunch, implying close proximity of the bunch
from an unstable fixed point of the 2:1 resonance.  
We should expect
diffusion to occur at relatively smaller modulation amplitudes. 
The resonance strength functions for $\Delta\phi_0\!=\!45^\circ$ 
do not differ much in value from those for 
$\Delta\phi_0\!=\!30^\circ$ in Fig.~\ref{strength-30deg-fig}.

\begin{figure}
\includegraphics[angle=-90,width=3.1in,viewport=75 10 560 686,clip]
{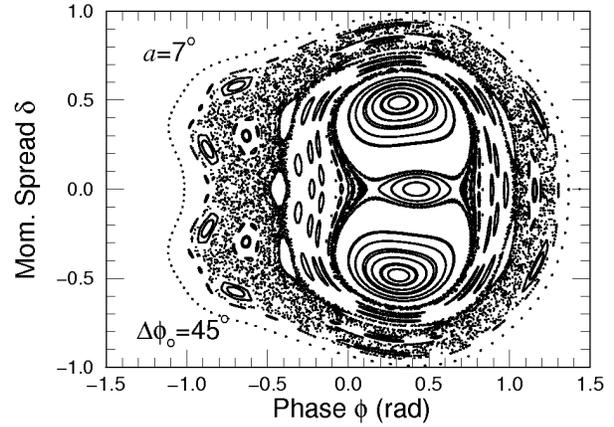}
\vspace{-0.1in}
\caption{\small 
Structure of phase space at $a=7^\circ$ and
$\Delta\phi_0=45^\circ$.   
\label{a7-45deg-8-fig}}
\vspace{-0.15in}
\end{figure}

\begin{figure}
\includegraphics[angle=-90,width=3.1in,viewport=75 10 560 686,clip]
{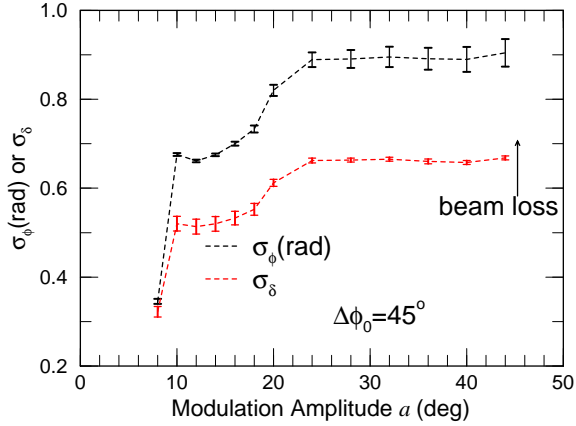}
\vspace{-0.1in}
\caption{\small (color)
Bunch spreads at various modulation amplitudes with $\Delta\phi_0=45^\circ$.
\label{bunchspread-45deg-fig}}
\vspace{-0.1in}
\end{figure}

The phase-space structure when
$\Delta\phi_0=45^\circ$ and $a=7^\circ$ is illustrated in 
Fig.~\ref{a7-45deg-8-fig}.
We first notice that the phase-space center is very close to the
2:1 resonance as speculated.  However, the tiny bunch there can only
spread out inside the thin stochastic layers of the 2:1 resonance, and
cannot reach the larger chaotic region between the island chains of the
8:5 and 7:3 resonances.
The bunch can diffuse into this region only when the chains of
higher-order islands enclosing the 2:1 resonance collapse.  This happens
when $a\approx9^\circ$, which 
explains the rapid jump of simulated bunch-spread results 
from $a=8^\circ$ to $10^\circ$ 
in Fig.~\ref{bunchspread-45deg-fig}.

\begin{figure}
\vspace{-0.025in}
\includegraphics[angle=-90,width=3.1in,viewport=75 10 560 683,clip]
{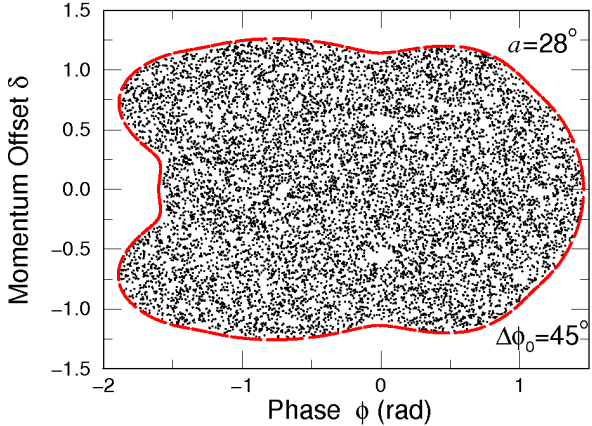}
\vspace{-0.13in}
\caption{\small (color)
Black: Bunch distribution 
at the last modulation period
after $\sim\!0.5$ million turns with  
$a\!=\!28^\circ$ and $\Delta\phi\!=\!45^\circ$.   
Red: Stroboscopic plot of one particle initially at $\phi=0.94$ rad
and $\delta=0$ at every modulation period.
\label{a14-45deg-82-fig}}
\vspace{-0.17in}
\end{figure}

\begin{figure}[t]
\includegraphics[angle=-90,width=3.1in,viewport=75 10 560 686,clip]
{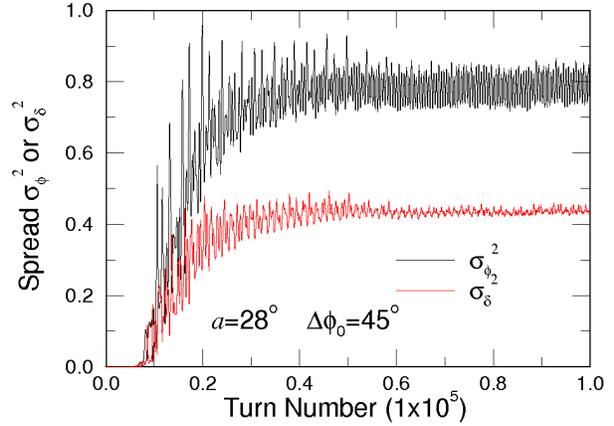}
\vspace{-0.1in}
\caption{\small (color) 
Longitudinal bunch-size squared 
turn by turn at $a\!=\!28^\circ$ and $\Delta\phi_0\!=\!45^\circ$
as diffusion develops.
\label{a28-45deg-93-fig}}
\vspace{-0.1in}
\end{figure}

\begin{figure}
\includegraphics[angle=-90,width=3.1in,viewport=75 10 560 683,clip]
{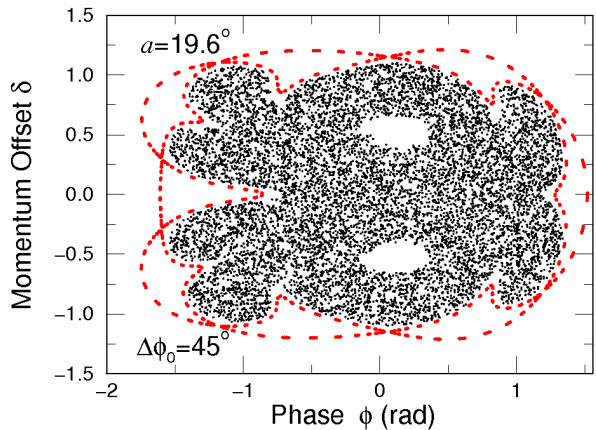}
\vspace{-0.13in}
\caption{\small (color)
Black: Bunch distribution at the last modulation period 
after $\sim\!0.5$ million turns
at $a=9.8^\circ$ and $\Delta\phi_0=45^\circ$.
Red: A chain of 8 islands 
forming a boundary for the diffused region.
At an increase to $a\approx20^\circ$, the separatrices of the island  chain
collapse, and the islands join the region of diffusion.
\label{a196-45deg-82-fig}}
\vspace{-0.19in}
\end{figure}

A typical diffused bunch distribution is shown in  Fig.~\ref{a14-45deg-82-fig},
corresponding to modulation amplitude
$a\!=\!28^\circ$  at the last modulation period
after roughly 0.5 million turns.
  Compared with
Fig.~\ref{a58-30deg-82-fig} at $\Delta\phi_0\!=\!30^\circ$, it is evident that
the chaotic area is much larger and the empty space inside is very much
smaller.  It also looks much more rectangular, and will provide a more
uniform linear density.  At the same time the modulation amplitude 
$a=28^\circ$ is about one-half smaller.
The red dots are stroboscopic loci of one particle initially located at
$\phi\!=\!1.46$~rad and $\delta\!=\!0$.  
These loci provide a well-behaved torus
bounding the diffused bunch.
The rms phase and momentum-offset spreads shown in 
Fig.~\ref{a28-45deg-93-fig} 
reveal  linear growths of $\sigma_\phi^2$ and $\sigma_\delta^2$,
demonstrating the occurrence of diffusion.
Compared with Fig.~\ref{a58-30deg-93-fig}
at $\Delta\phi_0\!=\!30^\circ$, equilibrium is reached much earlier.
This is understandable because the bunch initially is much closer to
the separatrices of the 2:1 resonance, and
obviously, will take less time to diffuse.
As illustrated in Fig.~\ref{bunchspread-45deg-fig}, there is another
jump of beam size around $a\approx 20^\circ$.
This can be explained by the hump of the synchrotron frequency around
action $J\approx0.7$ in Fig.~\ref{syntune-J-45deg-fig}. 
As a result, there are two sets of 8:3 resonances, 
one going out and one coming in as the modulation amplitude $a$ increases.  
The one going out has already broken at $a\approx7$.  The incoming set
encircling the chaotic region starts collapsing around $a=20^\circ$,
and the chaotic region is increased after that.
This is illustrated in \vspace{-0.1in}Fig.~\ref{a196-45deg-82-fig}.

\section{CONCLUSION\vspace{-0.03in}}

We have devised a method of phase modulation of the rf wave
to create
a large chaotic region in the central longitudinal phase space bounded
by well-behaved tori.  
To accomplish this, we require (1) 
large modulation amplitude so
that higher-order parametric resonances collapse to form a large
chaotic area, and (2) the initial position of the tiny
bunch inside 
this chaotic region.  Since the 
bunch is initially located
at the phase-space center, we must offset the relative phase of the
two-rf system so that the potential-well bottom is shifted 
away from the phase-space
center.  The maximum well-bottom offset and the corresponding
relative phase difference between the two rf's are \vspace{-0.07in}computed.

\end{document}